# First principle calculations with SIC correction of Fe-doped CuO compound

Fatima Zahra CHAFI[1*], Elmehdi SALMANI[2], Lahoucine BAHMAD[2], Najem HASSANAIN[1], Fares BOUBKER[1], and Ahmed MZERD[1]

[1]*Laboratory of Physics and Materials (LPM) - Mohammed V University, Faculty of Science- BP. 1014- Av. Ibn Batouta Rabat, Morocco*
[2]*Laboratory of Magnetism and Physics of High Energy (LMPHE-URAC-12) - Mohammed V University, Faculty of Science – BP. 1014- Av. Ibn Batouta Rabat, Morocco*

*Corresponding author e-mail address: chafifatimazahra@gmail.com*

## ABSTRACT

In this work, the electronic properties of Fe-doped CuO ($Cu_{1-x}Fe_xO$) thin films are studied by using a standard density functional theory (DFT) based on the ab-initio approach under the Korringa-Kohn-Rostoker coherent potential approximation (KKR-CPA). This study is carried out in the framework of the general gradient approximation (GGA) and self-interaction-corrected (SIC). The density-of-states (DOSs) in the energy diagram are illustrated and discussed. The computed electronic properties of our compounds confirm the half-metalicity nature of this material (CuO). In addition, the absorption spectra of the studied compound within the Generalized Gradient Approximation GGA, as proposed by Perdew–Burke–Ernzerhof (PBE) and GGA-PBE -SIC approximations are examined. Compared with the pure CuO, the Fermi-levels of doped structures are found to move to the higher energy directions. Finally, the effect of Fe-doping method in CuO can transform the material to half-metallic one, with a high wide impurity band in two cases of approximations local density approximation (LDA) and SIC method.

## 1. Introduction

The diluted magnetic semiconductors (DMS) are the basis of new materials. They have shown interesting results for new generation of spintronics devices. These materials are obtained by the substitution of a non-magnetic semiconductor by transition metal (TM) elements such as Co, Mn, Ni, Cr and Fe [1]. The DMSs showed a high Curie temperature $T_c$ and half-metallic behaviors. As well as other transition metal oxides, the Copper Oxide (CuO) material has anti-ferromagnetic ground state [2]. As a narrow band gap semiconductor, the CuO has been proved to be a good host material for diluted magnetic semiconductors [3, 4].

Recently, extensive studies have been carried out in transition elements-doped CuO on account of their potential uses in many technological fields such as: spintronics [5-7], memory devices [8], optoelectronics [9], gas sensors [10-13], solar cells [14] and more other applications [15-23].

To study the electronic properties of CuO, many theoretical techniques have been used like B3LYP (Becke three parameters, Lee, Yang and Parr) [24], PWPP-LSDA (Plane Wave Pseudo Potential-Local Spin Density Approximation) [25], CIPSI (Configuration Interaction by Perturbation Selected Iteratively) [26], CASSCF (Complete Active Space Self Consistent Field) [27], LSDA (Local Spin Density Approximation) [28] and PBE-GGA (Perdew functional with Generalized Gradient Approximation) [29] methods.

In this paper, we are interesting in the study of electronic properties of $Cu_{1-x}Fe_xO$ for specific concentration values (x=0.05, 0.10, 0.15), using the density function theory (DFT) under the Korringa-Kohn-Rostoker (KKR) method as well as the coherent potential approximation (CPA) with LDA (local density approximation) and SIC (self-interaction-corrected) approximations. This article is organized as follows: in next section we discuss the method and computational details. Section 3 illustrates a sketch of geometry of the studied compound. In section 4, we give results and discussion. Section 5 is devoted to a conclusion of this work.

## 2. Computational details

We use the Korringa–Kohn–Rostoker (KKR) method combined with the coherent potential approximation (CPA) within the local spin density approximation (LSDA) to study the electronic structure. This has the advantage of taking into account the randomness of the impurity elements. To parameterize the exchange energy we used the Generalized Gradient approximation (GGA) [30]. For more realistic description of the disordered local moments of the materials under investigation, the self-interaction corrected "SIC" approach is applied. This method is developed by Toyoda [31] within the KKR-CPA-SIC-GGA package as implemented into MACHIKANEYAMA2002 [32]. Due to hybridization of the orbital with the valance band and the energy gain upon localization of the orbital, the SIC approximation is found to be governed by the energy difference between the energy gain [33]. Compared to the GGA approximation, the SIC-GGA leads to an enhancement of the photoemission spectra. Free calculation of the SIC-GGA is similar to the GGA + U approximation [34]. The correction U is taken into account to fit the photoemission spectra in the dilute magnetic semiconductors (DMS). One of the most efficient methods is the CPA approximation which is considered to be one of the most used methods to study the band structure calculations. This method was developed by Akai and Dederichs [35], to investigate the transition metal alloys in As-based DMS [36].

## 3. Crystal Structure

Copper oxide (CuO) is commonly known as a Tenorite and p-type semiconductor with a band gap of 1.2 eV [37]. The crystal structure of CuO is monoclinic (see figure1) with C2/c symmetry, the dimension of unit cell is a= 4.69 Å, b=3.43 Å, c= 5.15 Å, α=ɤ=90° and β= 99.52°. The copper ions occupy positions ±(1/4, 1/4, 0 ; 3/4, 1/4, 1/2) along the lattice vectors, and the oxygen ions take positions ±(0, u, 1/4 ; 1/2, u+1/2, 1/4), where u = -0.584 is a parameter describing the relative positions of oxygen ions along this lattice.

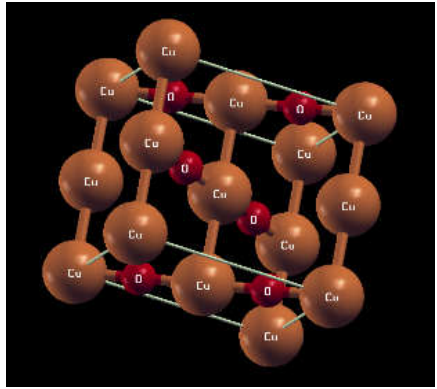

Figure 1. A sketch of the geometry of the monoclinic structure of pure CuO. The red circles represent Oxygen atoms and the brown ones represent the Copper atoms.

## 4. Results and discussion
### 3.1. Electronic structure

We generate the coherent potential approximation (CPA) from the atomic configuration of [Ar], for Cu $3d^{10} 4s^1$. The valence electrons used are 3d and 4s. In order to generate the coherent potential (CPA) approximation for O, we use [He] $2s^2 2p^4$. The valence electrons used are 2s and 2p. When using the density functional theory (DFT), it is difficult to model the transition metal oxides [38]. Since the local density (LDA) [39] and the generalized gradient approximations (GGAs) [40] do not consider the exchange and the correlation effects in transition metal oxides thus this leads to important self-interaction errors. For that reason, it is necessary to introduce the self-interaction correction (SIC) [41], and also the DFT+U [42] methods. On the other hand, the GW approximations (GWA) [43] give more precious results for the compound CuO [44-45]. The most used compound as a Mott or charge transfer insulator is the compound CuO in the energy band gap between 1.4 – 1.7 eV [46-49].

As known in the literature, the CuO shows a local magnetic moment of 0.65–0.69 μB [50], per formula unit. From figure 2(a), it is shown that there is no band gap in the GGA PBE approximation. But, when introducing the SIC approximation, this band gap arises as it is illustrated in figure 2(b). This is due to delocalization of electrons in such materials and nonzero self-interaction errors of DFT. When doping a semiconductor the Fermi level is displaced towards a high energy values. This is due to the fact that the doping action enhances the stability of the compound corresponding to the low energies. Furthermore we used the difference of energy: E-$E_f$. Our results are in good agreement with this statement, see figure 2.

On the other hand, the d-states cross the Fermi level with nonmagnetic states on the Cu atoms are illustrated in Figure 3. In addition, a direct band gap is shown in this figure at the Γ band reached at 1.5 eV. The magnetic moment is 0.67 μB as it is illustrated in figure 2(b). When exploring the experimental studies, the magnetic moment and band gap for CuO seem to be in good agreement with those found in the literature [46-50]. The transition from '*d*' level to '*t*' and '*e*' parts are separated by the crystal field effects. We illustrate in figure 4 the total and partial density of states (DOS) of the samples for the Fe-doping concentrations (5% and 10%). The results are obtained by the GGA PBE and the GGA PBE-SIC approximations (see Figure 4). Indeed, the effect of the crystal field environment is to split the atomic *3d* level into five-time degenerate *t* and *e* subgroups. The states $t_{2g}$ and $e_g$ are less localized in the valence band states. A near half-metallic behavior is found for the studied of Fe-doped CuO, according to the electronic structure calculations.

In order to highlight the inaccuracy of the GGA PBE, to describe the electronic structure, we introduce the SIC approximation with GGA PBE-SIC. The figure 4 illustrates the near half-metallic behavior in Fe, but no half-metallic behavior is shown in the compound $Cu_{1-x}Fe_xO$. This is in contradiction with the other investigation [51]. The *3d* levels of Fe-doped CuO are slightly shifted to a lower energy. This is due to the correction introduced by the SIC-GGA PBE calculations, see figure 4 for 5% and 10%. The valence band enhances the *p–d* hybridization according to the GGA-SIC approximation and localizes the *3d* impurities level in the bottom of this energy band. Moreover, increasing the Fe-doping effects is to enhance the band gap value from 1.5 eV to 3.1 eV.

The Table I resumes the calculated values of the local and total magnetic moments for different concentration values of the Fe-doped CuO within GGA and GGA-SIC approximations. The total magnetic moment is strongly located in the iron sites, while the local magnetic moment is found at the copper and oxygen atoms. Since the PBE-SIC is a correction of the used model, hence the obtained results are expected to be different than those obtained by the PBE approximation. Therefore, the difference found for the total moment without and within the SIC correction for the concentration 5% in table 1, is explained by the magnetic disorder related to the concept of a DLM state. However, two components of one magnetic atom are considered in opposition to the magnetization directions giving rise to the antiferromagnetic state. Otherwise, the other doping ratios 10 and 15 % correspond to the ferromagnetic case.

TABLE I. Total and local magnetic moment of $Cu_{1-x}Fe_xO$ thin films with GGA and GGA-SIC approximations

| Samples | PBE | | | | PBE-SIC | | | |
| --- | --- | --- | --- | --- | --- | --- | --- | --- |
| | Local moment with GGA | | | Total moment | Local moment with GGA | | | Total moment |
| | Cu | Fe | O | | Cu | Fe | O | |
| $Cu_{0.95}Fe_{0.05}O$ | 0.00448 | -3.15543 | -0.00835 | -0.3602 | -0.71773 | -3.81780 | -0.32432 | -2.2765 |
| $Cu_{0.90}Fe_{0.10}O$ | -0.00943 | 3.14476 | 0.01773 | 0.7195 | 0.70511 | -3.80297 | 0.22029 | 0.7601 |
| $Cu_{0.85}Fe_{0.15}O$ | -0.01153 | 3.13444 | 0.02771 | 1.0829 | -0.01062 | 3.1467 | 0.0286 | 1.1830 |

The magnetic energy difference ΔE between the FM and the DLM is calculated by:

$$\Delta E = E_{DLM} - E_{FM} \qquad (1)$$

This magnetic disorder is conveniently treated by the concept of a DLM state [52-53]. Two components of one magnetic atom are considered in the DLM states, in opposition to the magnetization directions.

As it is seen in table II, the calculated magnetic energy is compared with GGA and GGA-SIC approximations. From this table, it is found that the introduction of GGA-SIC approximation leads to important values of the energy differences compared to GGA without SIC approximations. Therefore, the critical temperature values are remarkably increased according to equation (2):

$$T_c = \frac{2}{3K_B} \frac{\Delta E}{C} \qquad (2)$$

Where C is the doping concentration and $K_B$ is the Boltzmann constant. The DLM state is described by $Cu_{1-x}Fe_{x/2\uparrow}Fe_{x/2\downarrow}O$ and the ferromagnetic state is described by $Cu_{1-x}Fe_{x\uparrow}O$.

The paramagnetic states of ferro-magnets are well described by the DLM states, see Ref. [54].

From the table II, it is well seen that the ferromagnetic state is the stable one for the Fe-doped CuO compound. Regarding the Katayama- Yoshida and Sato rule, the ΔE values are better described in the density of states.

The system exhibit a ferromagnetic state since its energy is so that: ΔE > 0 which corresponds to the ferromagnetic phase, see table II.

Within GGA-SIC Approximation, it is well known that the more positive larger ΔE corresponds to the more stable ferromagnetic state of the studied Fe doped CuO components.

The Curie temperature ($T_C$) is investigated from the total energy difference ΔE according to equation (2), in the framework of the Mean Field Approximation (MFA).

Table II summarizes the calculated values of $T_c$ by using MFA approximation. The square root of the concentration tendency is revealed. From this table, the increasing concentration values of iron leads to increasing values of $T_c$, in the PBE-SIC approximation. On the other hand, the PBE approximation without SIC correction reveals important values of $T_c$ for very low concentration values, as it is illustrated in table II.

TABLE II. The relative energies of FM and DLM orderings of $Cu_{1-x}Fe_xO$ thin films with GGA and GGA-SIC approximations.

| Samples | PBE | | PBE-SIC | |
|---|---|---|---|---|
| | ΔE ($E_{DLM}-E_{Ferr}$) (eV) | $T_C$ (°K) | ΔE($E_{DLM}-E_{Ferr}$) (eV) | $T_C$ (°K) |
| $Cu_{0.95}Fe_{0.05}O$ | 0.002321132 | 340.168 | 0.002585083 | 255.614 |
| $Cu_{0.90}Fe_{0.10}O$ | 0.003153801 | 487.939 | 0.003254483 | 503.516 |
| $Cu_{0.85}Fe_{0.15}O$ | 0.003289858 | 492.939 | 0.003361968 | 553.406 |

### 3.2. Absorption Spectra

Optical spectroscopy is another powerful method that allows the study of the electronic structure of crystals; this method has its advantages to complete the physical interpretation of the extracted information from the density-of-state (DOS). The absorption spectra of $Cu_{1-x}Fe_xO$ within GGA PBE and GGA PBE -SIC approximations are shown in Figure 5. It corresponds to transitions of 1s electrons of TM to empty states above the Fermi level. Meanwhile, other transitions are observed in this energy range as the transitions of 2s electrons. These transitions give only a very smooth contribution to the absorption spectrum while the absorption caused by transitions of 1s electrons changes very much near the absorption K-edge. That explains why the structure of the K-edge absorption spectra is determined by transitions of 1s electrons. Using the dipole approximation and taking into account these transitions, one can evaluate the absorption coefficients μ(E) [55-56]:

$$\mu(E) \propto |\langle \Psi_{1s}|e\nabla|\Psi_{4p}\rangle|^2 . n_{4p}(\hbar\omega - E_F + E_{1s}) \qquad (3)$$

Where: $\Psi_{1s}$ is the wave function of a 1s electron, $\Psi_{4p}$ is the wave function of a *4p* empty state, $n_{4p}$ is the density of *4p* states above the Fermi level, e is the light polarization.

When Fe atom is in a substitution position in the host semiconductor CuO, it is surrounded by six oxygen atoms located on the vertices of an octahedron environment. The octahedral crystal field of oxygen ions splits 3d-states of Fe, which are mainly located in the band gap, into $e_g$ and $t_{2g}$ levels with $e_g$ below $t_{2g}$ [57-59]. Exchange interactions further split these states into spin-up (↑) and spin-down (↓) states. Due to the tetrahedral arrangement of the ligands, the $t_{2g}$ -states of the Fe atoms hybridize with Fe 4p-states. The states resulting from hybridization of the deeper $t_{2g}\uparrow$- states show mainly Fe -3d character. From Figure 5, the outline of the main result and comparison is as follows: first the Fe-doped CuO K-edge within GGA PBE and GGA PBE -SIC exhibit a weak shoulder on a rising absorption curve that culminates in a strong peak. This shoulder is caused by the transition 1s to 4p that is authorized as an effect of the 4p-3d mixing due to the 4p-3d hybridization. In fact, the discrepancy between the results obtained between GGA and GGA-SIC results: in the first case, the correlation effects are not treated properly. In the second case, the localization of the 4p orbital of Fe in the valence band is overestimated. The shape of the spectra does not depend on the Fe content in CuO. Therefore, the Fe atoms have the same valence and local crystal structure in all the samples.

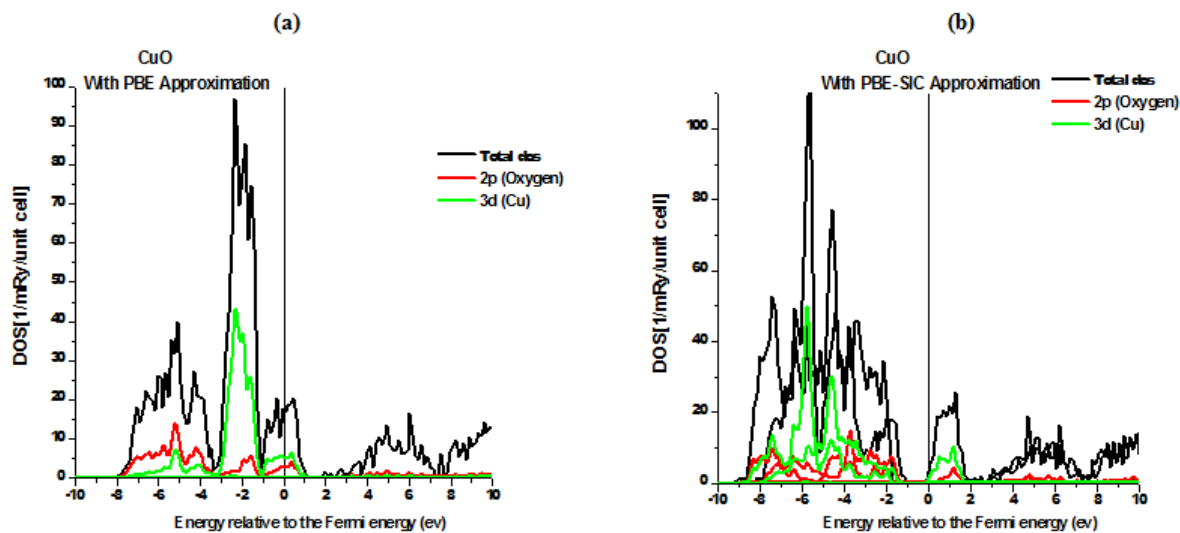

Figure 2.　The total projected DOS of CuO within GGA and GGA-SIC

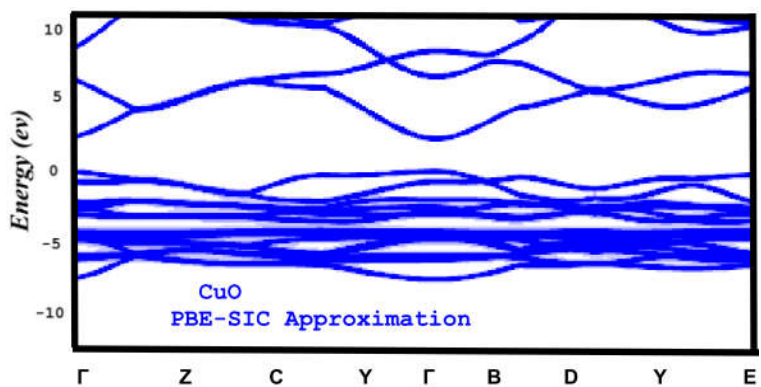

Figure 3.　The band structure of CuO within GGA-SIC

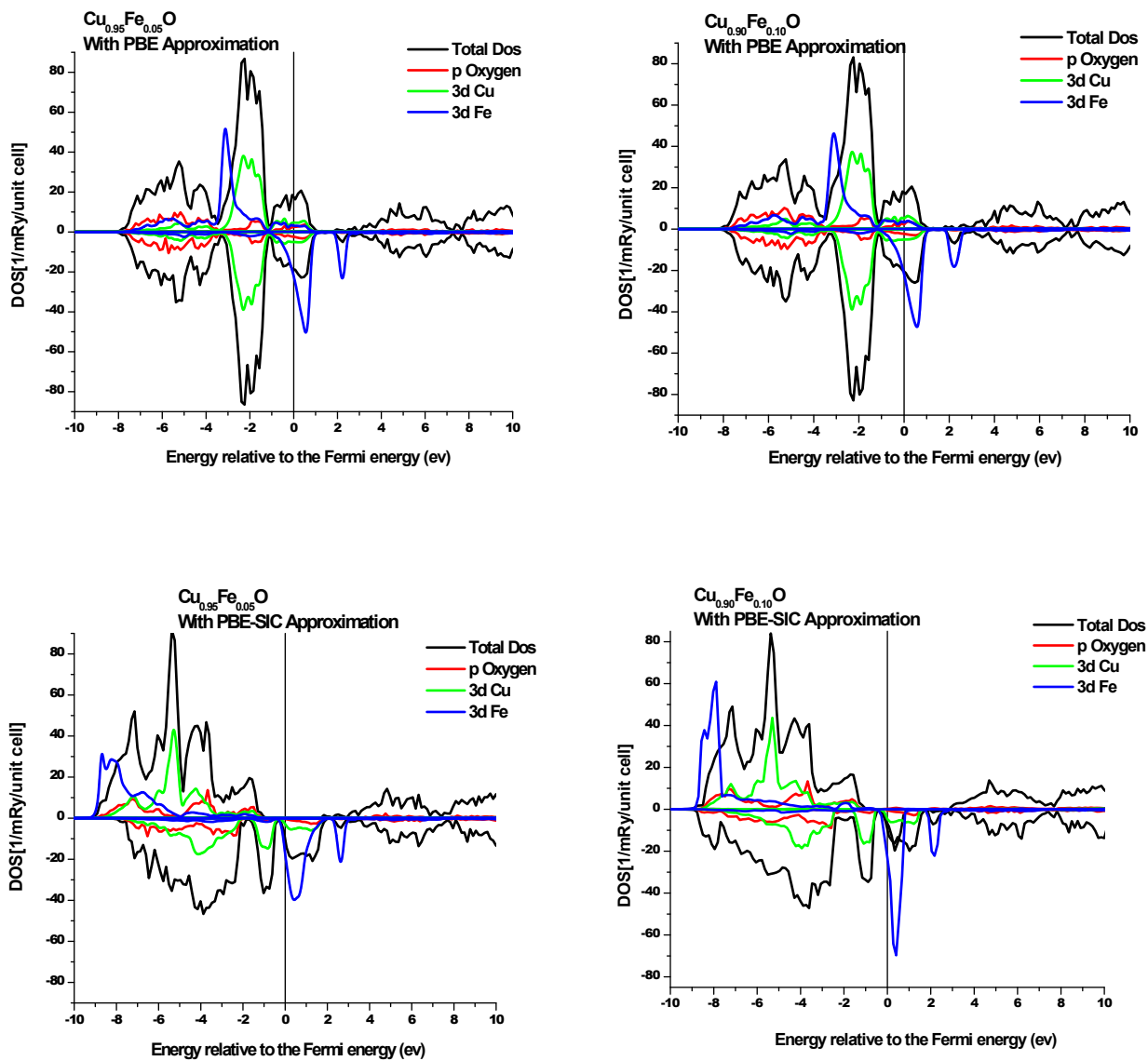

Figure 4. The total and d (TM) states projected DOS of CuO within GGA and GGA-SIC approximations

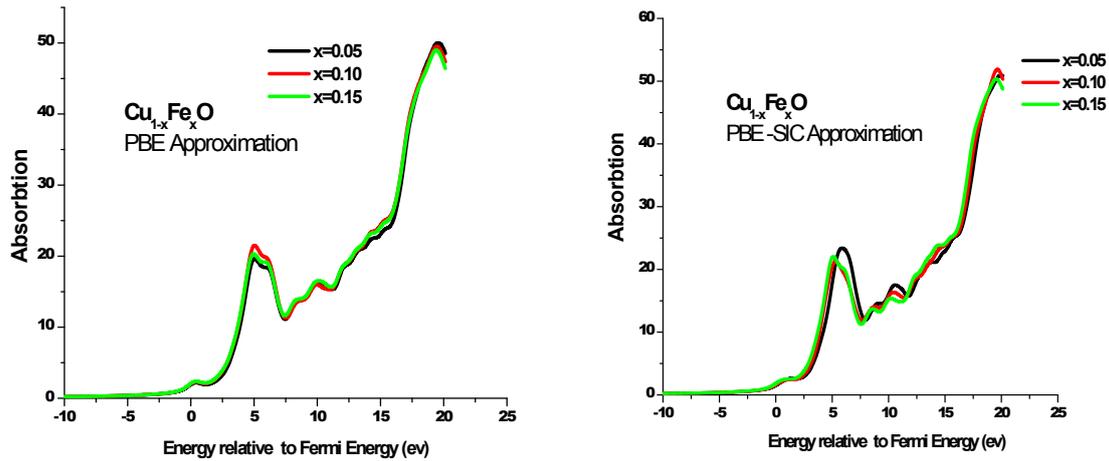

Figure 5. The calculated absorption spectrum at the K-edge CuO with GGA and GGA-SIC approximations

## 5. Conclusion

The electronic structures for $Cu_{1-x}Fe_xO$ have been investigated using the KKR-CPA method in connection with GGA without and within SIC corrections, respectively. The study of the electronic structure, the magnetic moment and absorption spectra of the investigated films with different concentration of iron Fe have been performed and discussed. Compared with the pure CuO, the Fermi-levels of the doped structures move to the higher energy directions. Finally, the increasing concentration of Fe-doping CuO can render the material half-metallic with a high wide impurity band in the framework of the two approximations LDA and SIC.